\documentclass[conference,a4paper]{IEEEtran}
\IEEEoverridecommandlockouts

\newenvironment{conditions}
{\par\vspace{\abovedisplayskip}\noindent\begin{tabular}{>{$}l<{$} @{${}={}$} l}}
	{\end{tabular}\par\vspace{\belowdisplayskip}}

\usepackage{cite}
\usepackage[cmex10]{amsmath}
\DeclareMathOperator{\arctantwo}{arctan2}
\usepackage[nolist,nohyperlinks]{acronym}
\usepackage{siunitx}
\usepackage{algorithmic}

\usepackage[pdftex]{graphicx}

\usepackage{subcaption}

\usepackage[obeyspaces]{url}

\usepackage{booktabs}

\begin{document}

\title{Towards Autonomous 1/8th Offroad RC Racing \\ The TruggySense Educational Platform}

\author{
	\IEEEauthorblockN{Robbe Elsermans (1), Jan Steckel (1 and 2)}
	\IEEEauthorblockA{(1) Faculty of Applied Engineering - Department of Electronics-ICT, University of Antwerp, Antwerp, Belgium,\\
		(2) Flanders Make Strategic Research Centre, Lommel, Belgium \\
		Email: robbe.elsermans@student.uantwerpen.be, jan.steckel@uantwerpen.be}

	\thanks{ }}

\maketitle

\begin{abstract}
This paper presents a state-of-the-art Data Acquisition System designed for off-road conditions, deployed on a Team Corally Kagama 1/8 Remote Controlled Vehicle. The system is intended to support Advanced Driver Assistance Systems in an educational context by providing valuable, consistent, and representative data. Key measurement systems are discussed to enable insights into the Remote Controlled Vehicles stability during and after off-road races. Furthermore, four experiments where conducted to evaluate the Data Acquisition Systems accuracy, stability, and consistency in replicating real-world vehicle behavior. The proposed Data Acquisition System platform serves as a solid foundation for use in engineering education, enabling integration with various Advanced Driver Assistance Systems algorithms to enhance vehicle control and overall performance, offering a new dimension to off-road racing. Additionally, real-time telemetry enables verification and validation of Advanced Driver Assistance Systems algorithms based on the live operating state of the Radio Controlled Vehicle during races.

\end{abstract}

\section{Introduction}

When driving an \ac{RCV} during a race, numerous external factors can influence its behavior. For example, each wheel provides specific traction suited to the track~\cite{VOSAHLIK2024105876, Christ03042021}, while the stiffness of the chassis, and the suspension system affects the \ac{RCV}s stability. Furthermore, the power output of the electric motor influences the actual steering angle due to Newton's first law of motion, in combination with the wheel slip and side-slip angle~\cite{10688791,10472888,newtonfirstlaw}.

%Indicating a gap
This raises the question: why is \ac{ADAS} not available for \ac{RCV} drivers during competition in off-road conditions?
%Purpose
One potential reason for this is that camera and lidar systems can drive an autonomous car fairly well with the proper advanced software attached to it~\cite{AVotE}. However, cameras and lidars are not used to provide stability to the vehicle in off-road situations such as difficult terrains where vehicle stability is a key factor~\cite{4282788, 10258400}. This raises a second question as such system could be an exciting platform for education and research in cyber-physical systems, including topics such as embedded systems, signal processing, and system control for increasing the stability of the vehicle itself. Nevertheless, it is embedded on real-scale vehicles, where it is not fully embedded on vehicles of magnitude 1/10 and 1/8~\cite{AVotE}. Furthermore, \ac{ADAS} for \ac{RCV}s could create a competition as these aid techniques can provide drivers an advantage in off-road terrains. Therefore, it is able to open a new era in \ac{RCV} in which students and researchers can engage in competition.

%How the paper is structured
In this paper, a first step towards such a framework is deployed where the main focus is within creating a stable and representative \ac{DAQ}. Moreover, the \ac{DAQ} can be utilized as ground truth for the \ac{ADAS} system such that it can rely on the data acquired from the \ac{DAQ}. Furthermore, the developed system can serve as a higher educational purpose where students and researches can utilize it as foundation. First, an overview of the used hardware is given, Next, the experimental setups and results are presented. Finally, a discussion of these results and shortcomings of the system combined with a conclusion that holds future work.

\section{Hardware Architecture}
\begin{figure*}[]
	\begin{subfigure}{0.32\textwidth}
		\centering
		\includegraphics[width=1\textwidth]{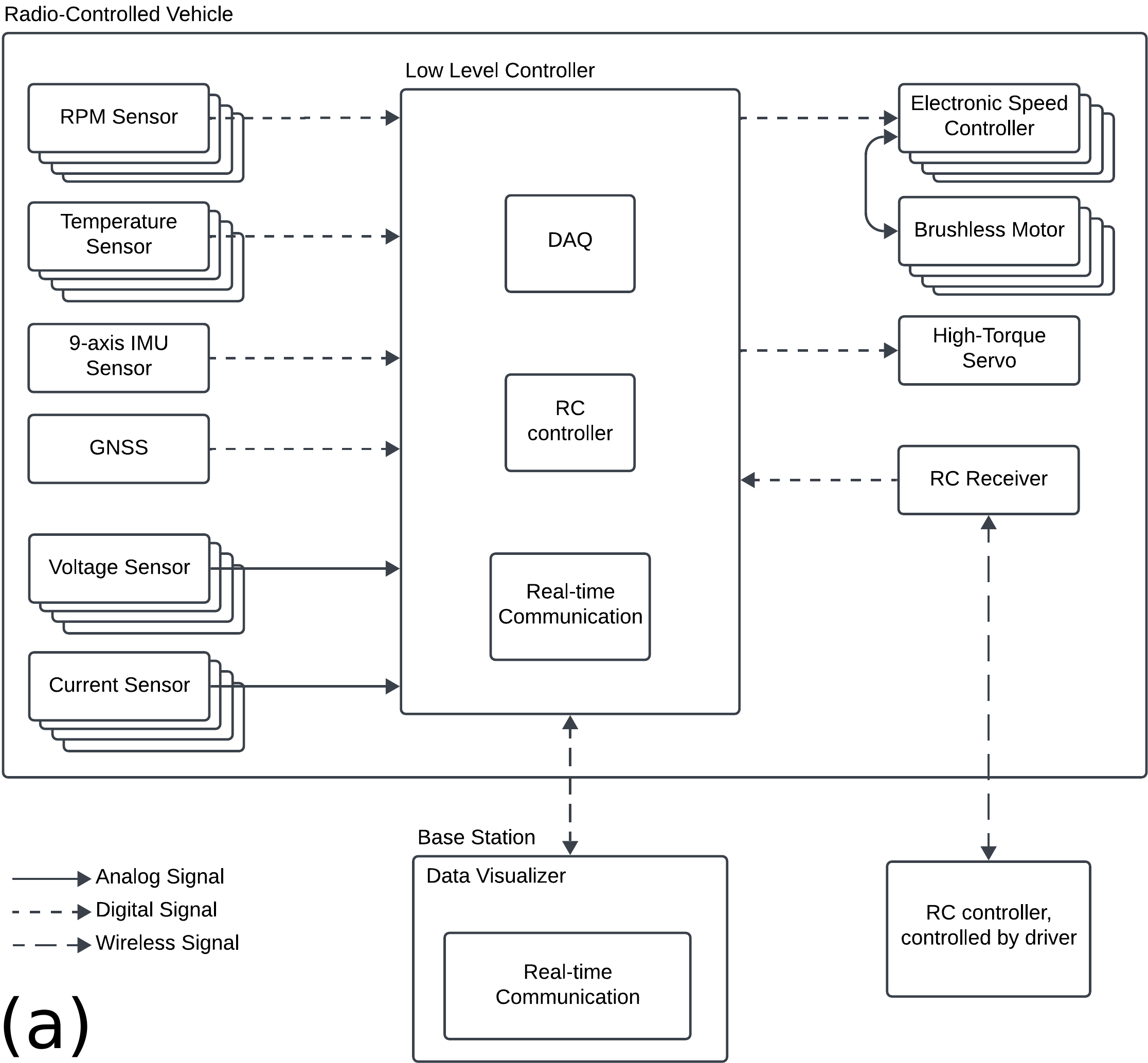}
	\end{subfigure}
	\begin{subfigure}{0.32\textwidth}
		\centering
		\includegraphics[width=1\textwidth]{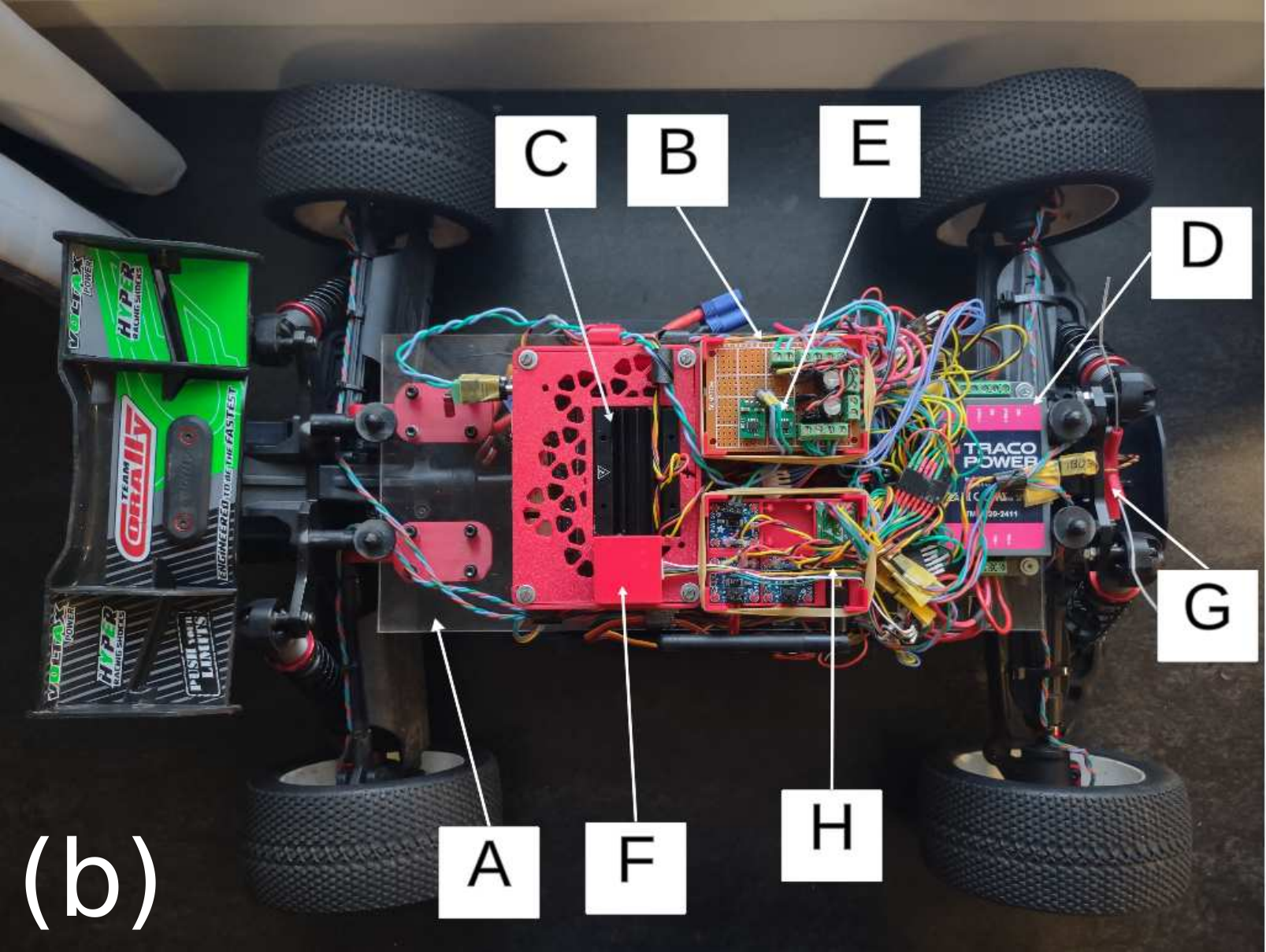}
	\end{subfigure}
	\begin{subfigure}{0.32\textwidth}
		\centering
		\includegraphics[width=1\textwidth]{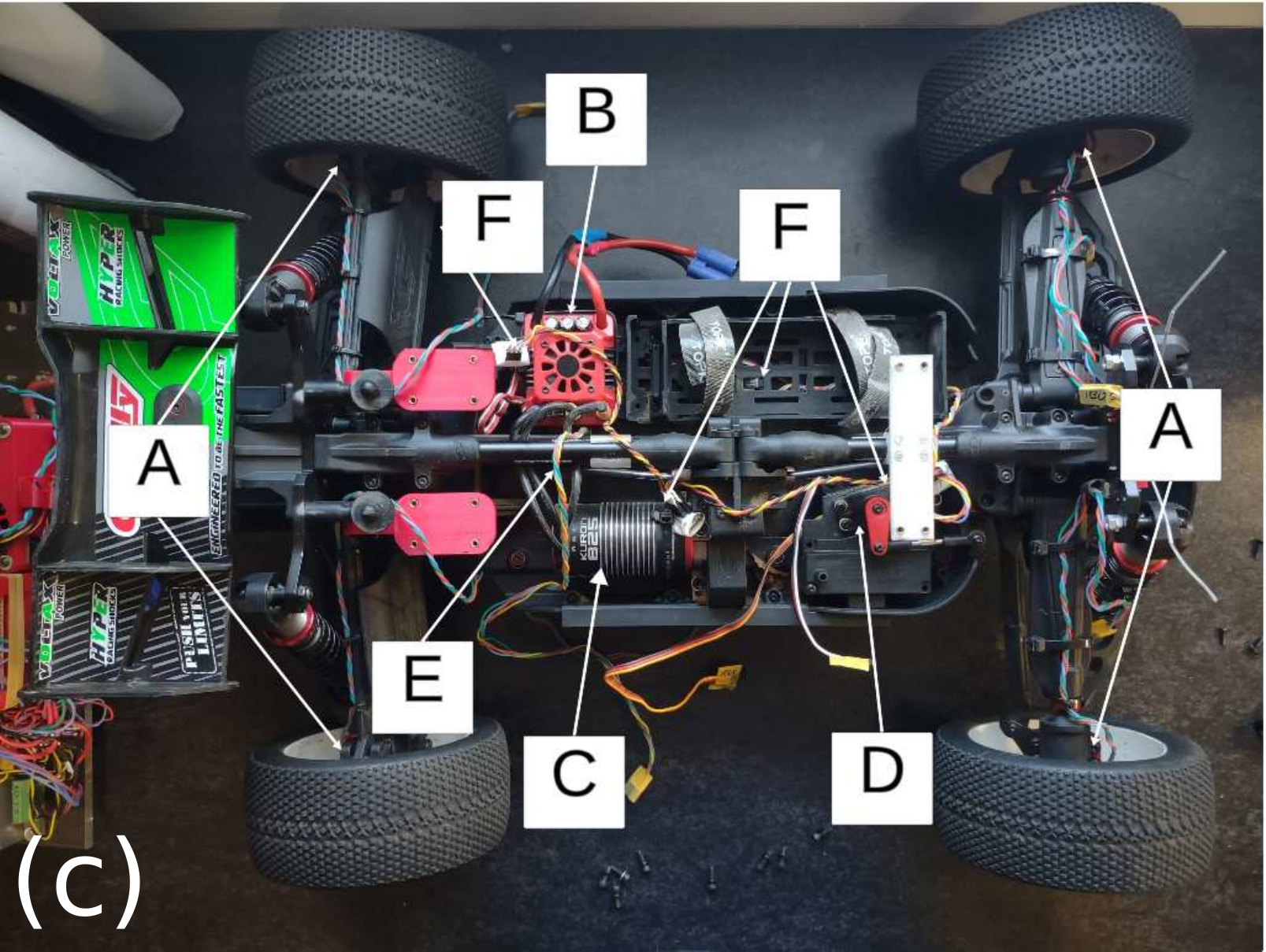}
	\end{subfigure}
	\caption{
		(a) representing the abstract block diagram that covers the entire system of the framework. Here, the \ac{LLC} is our \ac{DAQ} which will capture various sensors such as the GPS, encoders for all wheels and motors, \ac{IMU}, etc. Additionally, the \ac{LLC} is responsible for actuating components such as the \ac{ESC}s and the steering servo of the \ac{RCV}, denoted as the \textit{RC-Controller} block. Lastly the \textit{Real-time Communication} module providing telemetry to the base station. (b) representing the hardware layout which is used on the Corally Kagama chassis where 
		(b).a is the base plate used for mounting various modules, 
		(b).b the \ac{LLC}, 
		(b).c the \ac{HLC}, 
		(b).d a 5.1~\si{\volt} 4~\si{\ampere} DC-DC step-down converter, 
		(b).e a 3.3~\si{\volt} 1~\si{\ampere} DC-DC step-down converter, 
		(b).f the GPS module, and
		(b).g X6B RF receiver
		(b).h NRF24 telemetry module.
		(c) depicts the chassis in its bare essence without the base plate mounted.
		(c).a the wheel encoders,
		(c).b the \ac{ESC} to actuate the used brusheless motor,
		(c).c the brusheless motor,
		(c).d the steering servo,
		(c).e the \ac{IMU}, and 
		(c).f the temperature sensors.
	}
\label{fig:setup}
\end{figure*}

In this work, a specific existing \ac{RCV} platform is utilized which is the Team Corally Kagama 1/8. Onto this platform, various sensors needs to be installed in different areas. These sensors are managed by a \ac{LLC} that can reliably acquire sensor data at high-speeds. Additionally, the \ac{LLC} requires the ability to control the \ac{RCV} and provide telemetry to a remote unit as depicted in Fig.~\ref{fig:setup}.a. The Team Corally Kagama 1/8 \ac{RCV} platform is not built to be utilized in such way. Therefore, a small study is performed on how and where the sensors should be placed on the vehicle in order to not degrade the structural integrity and stability. Furthermore, the added sensors and modules need some sort of protection from the harsh environment such as rocks and sand flying around. Therefore, the vehicles top cap needs to be mountable after the modifications and additions to the \ac{RCV}.

In Fig.~\ref{fig:setup}b, the system layout is shown. The base plate is shown in Fig.~\ref{fig:setup}b.a which is designed to serve as a mounting point for all additional modules and components. These include the \ac{HLC} shown in Fig.~\ref{fig:setup}b.c, a DC-DC 5~\si{\volt}, 4~\si{\ampere} step-down power supply in Fig.~\ref{fig:setup}b.d, a DC-DC 3.3~\si{\volt}, 1~\si{\ampere} step-down converter in Fig.~\ref{fig:setup}b.e, the \ac{GPS} module in Fig.~\ref{fig:setup}b.f, and additional circuits designed to enhance the safety and performance of the \ac{DAQ} system shown in Fig.~\ref{fig:setup}b.b. The telemetry module in Fig.~\ref{fig:setup}b.h. Lastly, the transceiver module for the remote controller depicted in Fig.~\ref{fig:setup}b.g. As mentioned previously, the platform used is an existing Team Corally Kagama 1/8 \ac{RCV}, depicted in Fig.~\ref{fig:setup}c. On this platform, the \ac{ESC} is shown in Fig.~\ref{fig:setup}c.b, and the brushless motor is illustrated in Fig.~\ref{fig:setup}c.c. The steering servo can be seen in Fig.~\ref{fig:setup}c.d. Additional components have been integrated, such as the four encoders presented in Fig.~\ref{fig:setup}c.a, the temperature sensors shown in Fig.~\ref{fig:setup}c.f, and the \ac{IMU} located at the near center of the \ac{RCV} pictured in Fig.~\ref{fig:setup}c.e.

\subsection{Wheel Rotational Speed}

\begin{figure}
		\centering
		\includegraphics[width=1\linewidth]{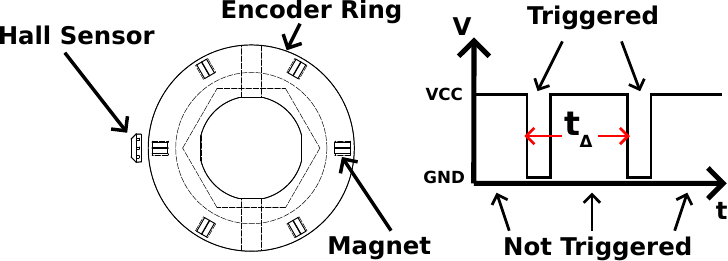}
		\caption{The used encoder ring structure where $X$ neodymium permanent magnets are utilized to determine the wheel speed. Here, six magnets are presented whereas in the used platform, 14 magnets are utilized.}
		\label{fig:encoder}
\end{figure}

The wheel and motor rotational speed can give insights into wheel slip and stability of the whole \ac{RCV} with respect to the off-road surface \cite{10750301}. Therefore, measuring the rotational speed is essential for each wheel and motor. To archive this, Hall effect sensors are used which generate a voltage proportional to the strength of the magnetic flux $\phi$ in one direction \cite{ramsden2011hall}. To determine the rotational speed, permanent magnets are used in conjunction with a fixed Hall effect sensor depicted in Fig. \ref{fig:encoder}. A specialized mount is designed to hold $14$ magnets. The number of magnets affects the \ac{PPR}. More magnets provide higher resolution but introduces faster pulses, while fewer magnets result in lower resolution and reduce the pulses. 

In order to determine the rotational speed at high speeds with minimal delay, the difference between two pulses $t_\Delta$ produced by the Hall effect sensor can be used as described in equation \ref{eq:difference_pulse_eq} . 
Furthermore, this difference can be utilized to determine the frequency of the pulses $f_{p}$ such as in equation \ref{eq:frequency_pulse_eq}. Additionally, the radius of the wheel $r$ in combination with the angular velocity $\omega_{w}$, denoted in equation \ref{eq:omega_wheel_eq}, can be used to determine the wheel speed $v_{w}$ such as in equation \ref{eq:wheel_speed_eq}.

\begin{align}
	t_\Delta = t_2 - t_1 ~(\si{\second})
	\label{eq:difference_pulse_eq}
\\
	f_{p} = \frac{1}{t_\Delta} ~(\si{\hertz})
	\label{eq:frequency_pulse_eq}
\\
	f_{w} = \frac{f_{p}}{X} ~(\si{\hertz})
	\label{eq:frequency_wheel_eq}
\\
	\omega_{w} = {f_{w}} \cdot 60 ~(\unit{rev/\minute})
	\label{eq:omega_wheel_eq}
\\
	\omega_{w} = \frac{n_{p}}{X \cdot t_\Delta} \cdot 60 ~(\unit{\kilo\meter/\hour})
	\label{eq:wheel_speed_eq_V2}
\\
	v = \frac{d\cdot\pi\cdot \omega_{w} \cdot 60}{1000} ~(\unit{\kilo\meter/\hour})
	\label{eq:wheel_speed_eq}
\end{align}

Where, $t_\Delta$ represents the time interval between two pulses denoted in seconds, $f_p$ defines the pulse frequency in Hertz, and $X$ describes the number of magnets installed on the encoder ring. Furthermore wheel's rotational speed is given by $\omega_{w}$ in revolutions per second, its diameter by $d$ denoted as meter, and its linear velocity by $v$ in kilometer per hour. Additionally, $n_p$ indicates the number of pulses detected over a specified time period. When a vehicle is not in motion, the \ac{DAQ} will not receive any pulses. Therefore, another approach is utilized where a task will run at a period $t_\Delta$. During this inactivity, the pulses triggered by an encoder will be stored in a counter $n_{p}$. When the task fires at time-point $t_\Delta$, the $n_{p}$ is used to calculate the rotational speed based on the used magnets $X$ such as Equation \ref{eq:wheel_speed_eq_V2}.

\subsection{Acceleration, Velocity and Direction}
The sensor proposed in this framework to enable Acceleration, Velocity and Direction of the vehicle is the BNO085, manufactured by CEVA Hillcrest Laboratories. This sensor is a nine-degree-of-freedom \ac{IMU}, which include a gyroscope, an accelerometer, and a magnetometer. As the names suggest, the gyroscope measures angular velocity, providing information about orientation. Moreover, the accelerometer detects acceleration and vibration within a structure whereas the magnetometer monitors changes in Earth's magnetic field. All three of these sensing units are combined in the BNO085. Even more, the BNO085 has its own computational unit, which is embedded within the sensor package, thus enabling onboard processing of sensor data.

\begin{figure}
	\centering
	\includegraphics[width=0.8\linewidth,trim={0 1cm 0 0.5cm},clip]{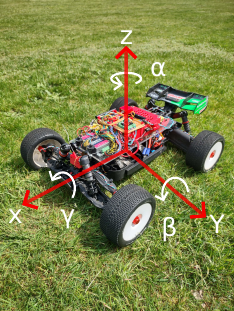}
	\caption{The Euler angles expressed as yaw $\alpha$, pitch $\beta$, and roll $\gamma$ on the Team Corally Truggy 1/8 \ac{RCV} platform on the XYZ-axis of the \ac{RCV}.}
	\label{fig:rollpitchyaw}
\end{figure}

In this framework, the Euler angles, Quaternions, and acceleration are determined by the \ac{LLC}. Quaternions offer the advantage of avoiding the issues associated with three-dimensional representations, such as those encountered with Euler angles, and eliminate the possibility of \textit{Gimbal Lock}. \textit{Gimbal Lock} refers to a phenomenon in which two Euler angles become aligned, resulting in the loss of one-degree-of-freedom \cite{10903617}. Nevertheless, the choice between quaternions and Euler angles for motion representation ultimately depends on the specific algorithm and the user's preference. Quaternions are defined in four dimensions, as outlined in equation \ref{eq:quaternion_base_equation}, and their general form is given by equation \ref{eq:quaternion_equation}. In contrast, Euler angles are a three-dimensional representation characterized by the parameters of yaw $\alpha$, pitch $\beta$, and roll $\gamma$ depicted in Fig.~\ref{fig:rollpitchyaw}. These can be determined through quaternions as denoted in equation \ref{eq:quaternion_to_euler}.

\begin{align}
	a + bi +cj + dk
	\label{eq:quaternion_base_equation} \\
	i^2=j^2=k^2=ijk = -1 
	\label{eq:quaternion_equation}
\end{align}
\begin{conditions}
	a, b, c, d 	& real numbers ${\rm I\!R}$ \\
	i, j, k		& the basis vectors 
\end{conditions}

The BNO085 provides quaternion motion data but not an Euler representation in \ac{I2C} operation mode. Therefore, the quaternion data is used to determine the Euler angle in degrees based on equation \ref{eq:quaternion_to_euler}. 

\begin{equation}
\begin{split}
	\alpha = \arctantwo  \left( \frac{2 \cdot ( b \cdot c + d \cdot a )}{a+b-c-d} \right) \cdot \left( \frac{360}{2\pi}\right) ~(\si{\degree}) \\
	\beta = \arcsin   \left( \frac{2 \cdot ( b \cdot d + c \cdot a )}{a+b+c+d} \right)\cdot \left( \frac{360}{2\pi}\right) ~(\si{\degree}) \\
	\gamma = \arctantwo \left( \frac{2 \cdot ( c \cdot d + b \cdot a )}{a-b-c+d} \right)\cdot \left( \frac{360}{2\pi}\right) ~(\si{\degree}) \\
	\label{eq:quaternion_to_euler}
\end{split}
\end{equation}

\begin{conditions}
	\arctantwo 			& the 360\si{\degree} version of the ${\arctan}$ \\
	 \alpha, \beta, \gamma	& the Euler angles
\end{conditions}

The \ac{LLC} determines the acceleration through the \ac{IMU} that is equipped with its own computational unit which can be utilized by the \ac{LLC} to simplify the code. In this framework, the \ac{LLC} reads precomputed acceleration values in the x, y, and z directions of the \ac{RCV} presented in Fig.~\ref{fig:rollpitchyaw}.

\subsection{Current and Voltage Measurement}
This framework operates at three voltage levels namely 3.3~\si{\volt}, 5.1~\si{\volt}, and 16.8~\si{\volt}. The 3.3~\si{\volt} and 5.1~\si{\volt} rails are derived from the main 16.8~\si{\volt} supply provided by a 4-cell Li-Po battery. To monitor system performance and detect anomalies, both the current consumption and output voltage of each voltage branch are tracked.
Furthermore, the Li-Po battery depletes over time whereas in loading stage, this depletion can be very fast. Moreover, it is essential to monitor its state to avoid overloading or deep discharging, which could damage the battery or the system. During racing conditions, current demand can peak at a maximum of 300~\si{\ampere}, with the selected 4-cell, 60~C, 5000~\si{\milli\ampere\hour} Li-Po battery. Measuring such high currents with a series resistor is impractical due to the resulting voltage drop and power loss, which would degrade vehicle performance. Therefore, a Hall-effect current sensor is used instead. This sensor detects the magnetic field generated by the current-carrying conductor, making it well-suited for high-current measurements. Although, it lacks precision below 5~\si{\ampere}.
In order to measure the low currents on the platform such as the current consumption of the \ac{LLC} and \ac{HLC}, a series resistor is employed. Here, the current is calculated based on the voltage drop across the known resistance.

\subsection{Global Position}
Driving the \ac{RCV} on an off-road and harsh track, presents a particularly demanding challenge due to the unpredictable nature of the terrain and the way in which it evolves over time. The optimal driving lines or track patterns are not static, it can shift significantly after just a few laps due to the fact that multiple competitors are navigating the same track patterns. Each vehicle contributes to the gradual alteration of the track surface, potentially creating fresh grooves, uneven surfaces, or scattered debris that make previously efficient lines less viable or even dangerous.
Therefore, the specific path chosen by the driver throughout the course becomes a valuable source of data. By analyzing the trajectories taken during each lap, it is possible to identify trends in driver behavior and determine which paths are most frequently selected or prove to be the most effective under evolving conditions \cite{AVotE}. This information can be vital in refining driving strategies and making informed decisions in future races. Furthermore, these insights can contribute to the identification and prediction of optimal racing lines, allowing teams to adapt to the changing terrain and enhance overall performance based on empirical positioning data gathered from previous runs \cite{10258400}.

\subsection{Temperature}
When testing a new \ac{ADAS} algorithm within the framework, certain flaws may cause sudden twitches or sharp actuation transitions. These abrupt movements can overheat critical components such as the \ac{ESC}, brushless motor, and steering servo. Although Li-Po batteries can handle current spikes, excessive heating can make vehicle operation unsafe. Therefore, monitoring these components is vital to detect issues early and prevent catastrophic failure.

The framework uses DS18B20+ temperature sensors with the OneWire protocol. Each sensor has a unique identifier that gives the possibility to configure these sensors in a bus configuration that reduces wiring to the \ac{LLC}. On the one hand, this setup simplifies hardware where on the other hand, it increases the time needed to read temperature data. Furthermore, these sensors track ambient temperature where low sampling rates are sufficient. Moreover, high precision is unnecessary as a 0.1~\si{\degreeCelsius} change is negligible, whereas a 10~\si{\degreeCelsius} shift may indicate a critical problem.

\subsection{Data Transmission}

To provide the driver with feedback during experiments, data can be stored offline and reviewed afterward. However, for improved usability and real-time validation of deployed \ac{ADAS} algorithms and the overall condition of the \ac{RCV}, real-time monitoring is essential. This framework establishes a wireless connection between the \ac{RCV} and its driver using the E01-ML01DP5 module, operating in the 2.4~\si{\giga\hertz} band \cite{10810938}. Furthermore, it includes a power amplifier up to 20~\si{\decibel}m transmit power and a 10~\si{\decibel} gain in receiver sensitivity which gives a theoretical range up to 2.5~\si{\kilo\meter}. Moreover, data rates range from 250~\si{\kilo\bit\per\second} to 2~\si{\mega\bit\per\second} whereas higher rates reducing transmission range.

The E01-ML01DP5 module supports a maximum payload size of 32 bytes where this framework exhibits a data structure of 228 bytes. To handle this, data is split into 30-byte chunks, reserving the last 2 bytes for a chunk counter. Furthermore, this counter allows the receiver at the base station to reassemble the data in the correct order. Although the E01-ML01DP5 module supports in-sequence packet transmission, these identifiers provide redundancy and help identify lost packets.

\section{Experimental Setup and Results}

\begin{figure}
	\centering
	\includegraphics[width=1\linewidth]{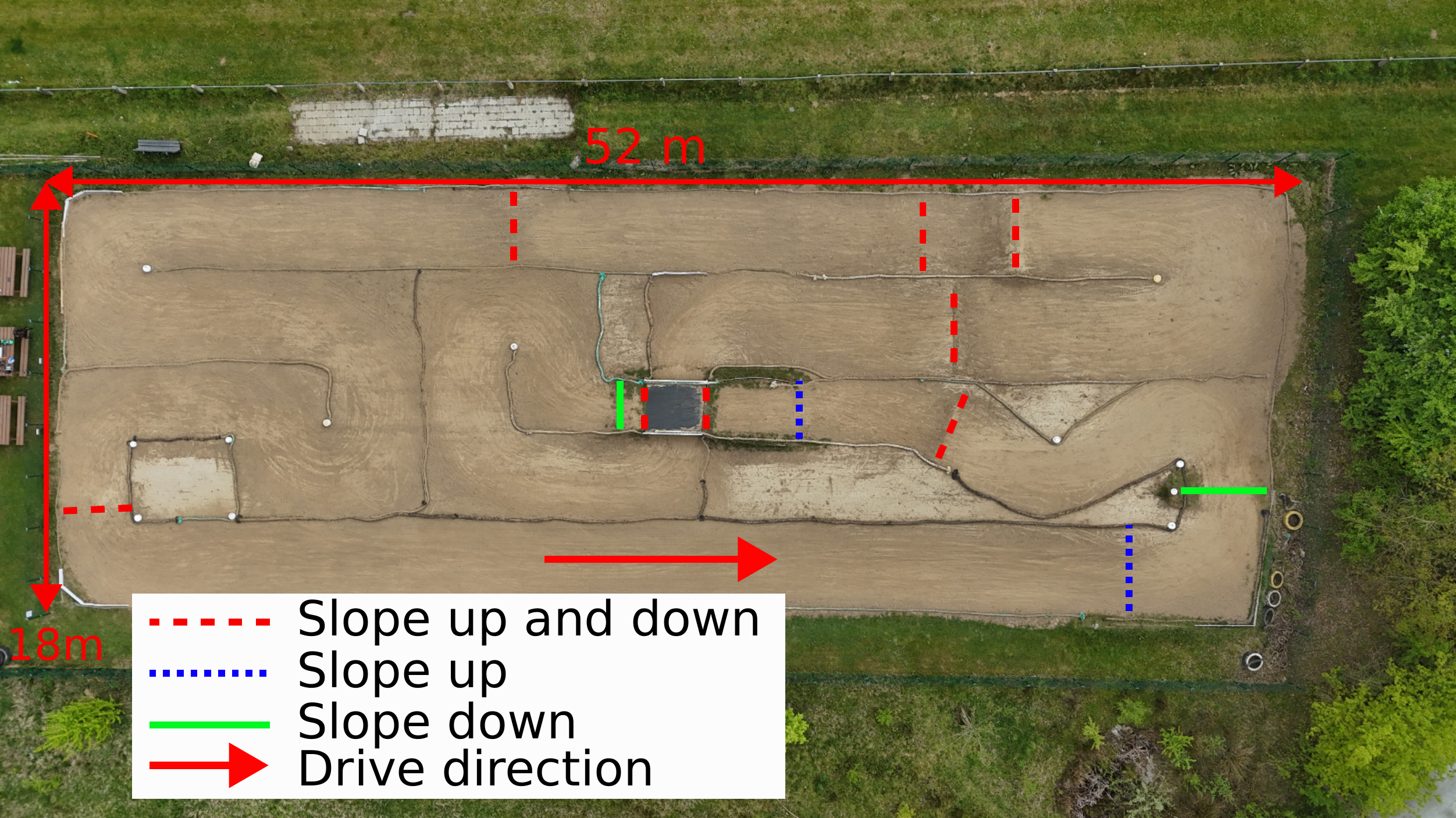}
	\caption{The used test trajectory maintained by RC Racing Fun Bornem located at Bornem, Antwerp, Belgium. There are different terrain features embedded in the track such as hills, uneven terrain, a tunnel, sharp corners, and straight paths, making this track a challenge to drive. Therefore, this track is suited for the test environment as different elements can be tested such as traction on the road, suspension system, stability with respect to cornerings, etc. This track has a one way direction denoted by the arrow. This drive direction is always used in each race.}
	\label{fig:predefined_path}
\end{figure}

\begin{figure*}[t]
	\centering
	\begin{subfigure}{.49\textwidth}
		\centering
		\includegraphics[width=1\linewidth]{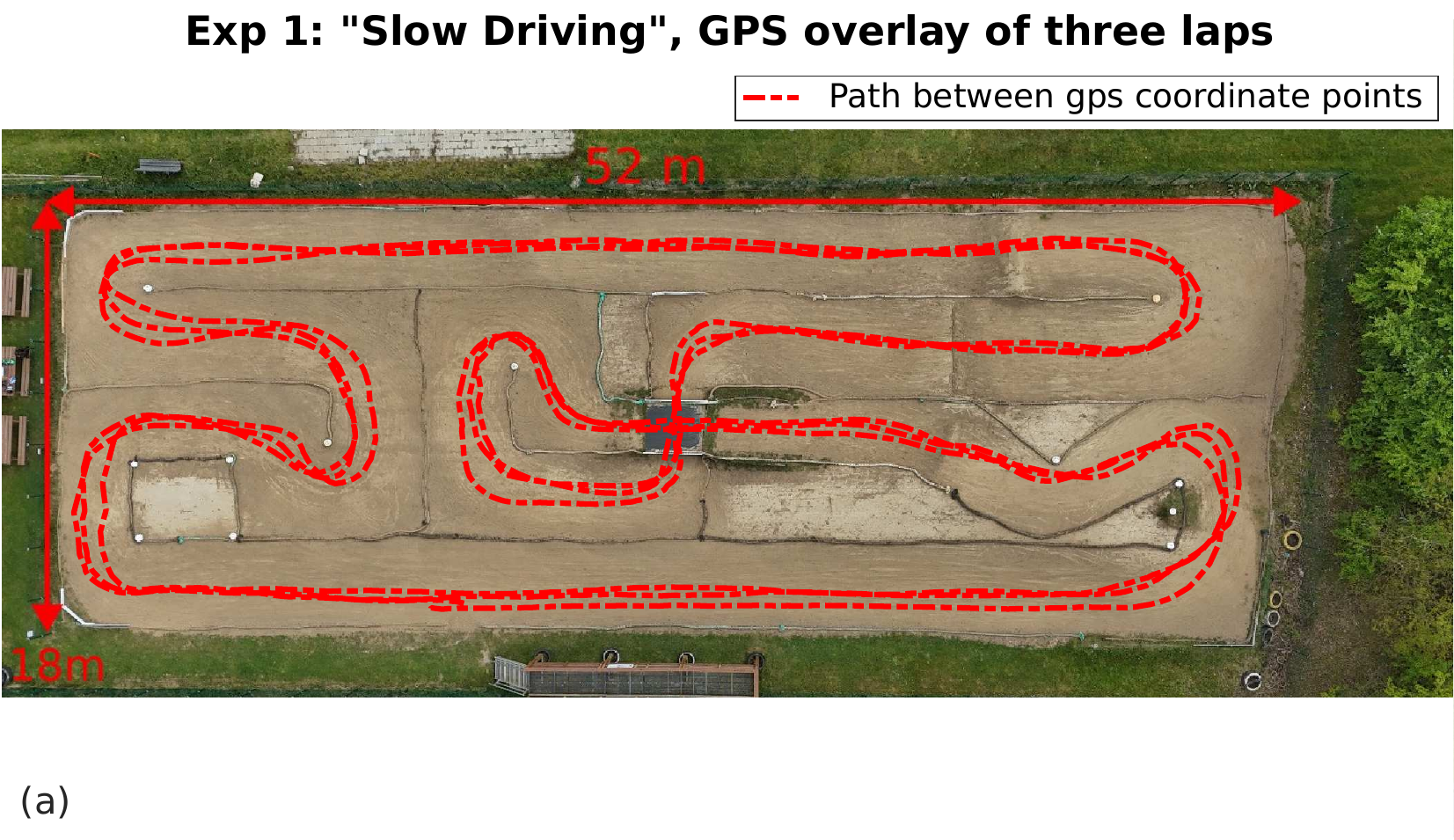}
	\end{subfigure}
	\begin{subfigure}{.49\textwidth}
		\centering
		\includegraphics[width=1\linewidth]{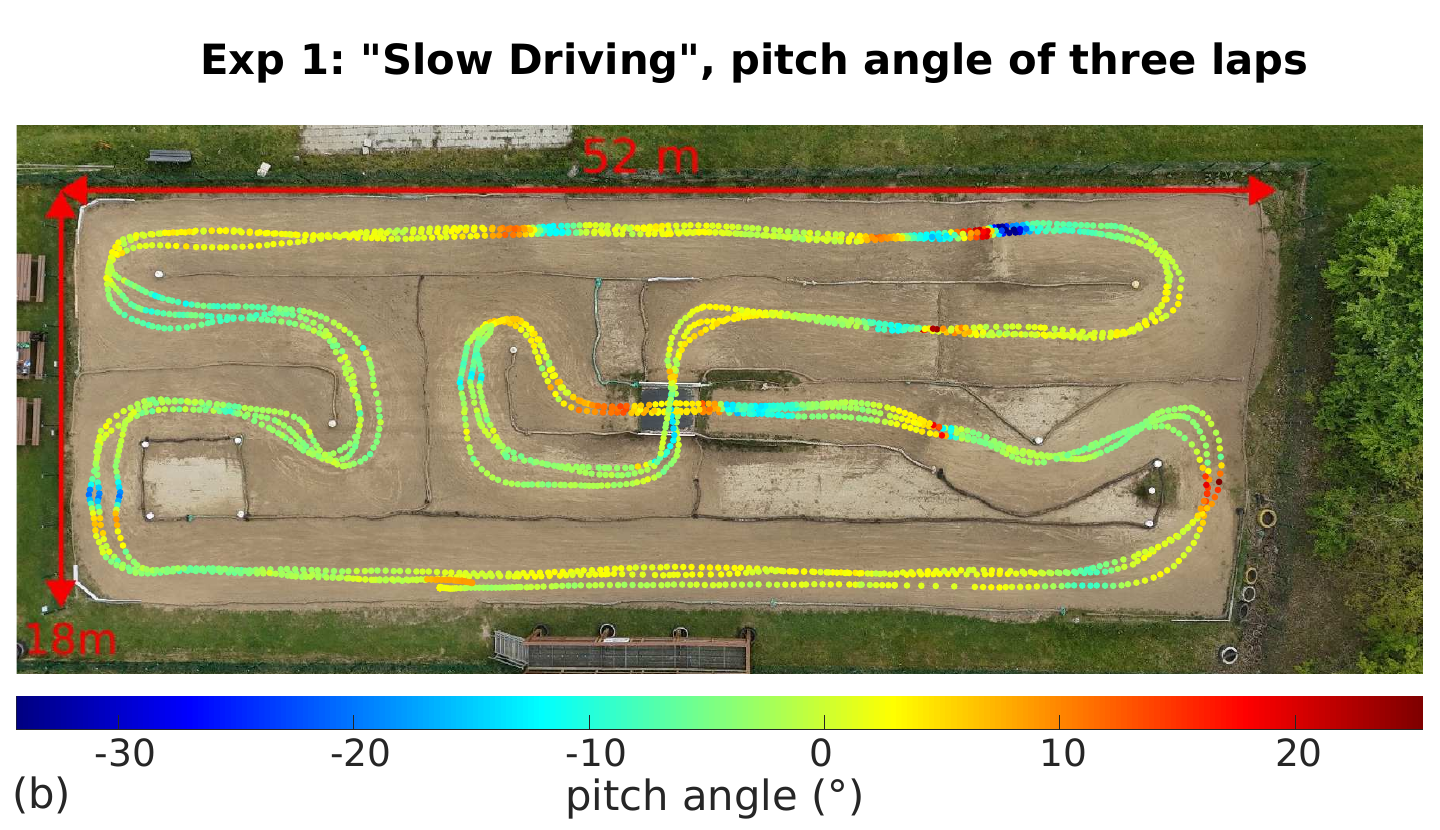}
	\end{subfigure}
	\begin{subfigure}{.49\textwidth}
		\centering
		\includegraphics[width=1\linewidth]{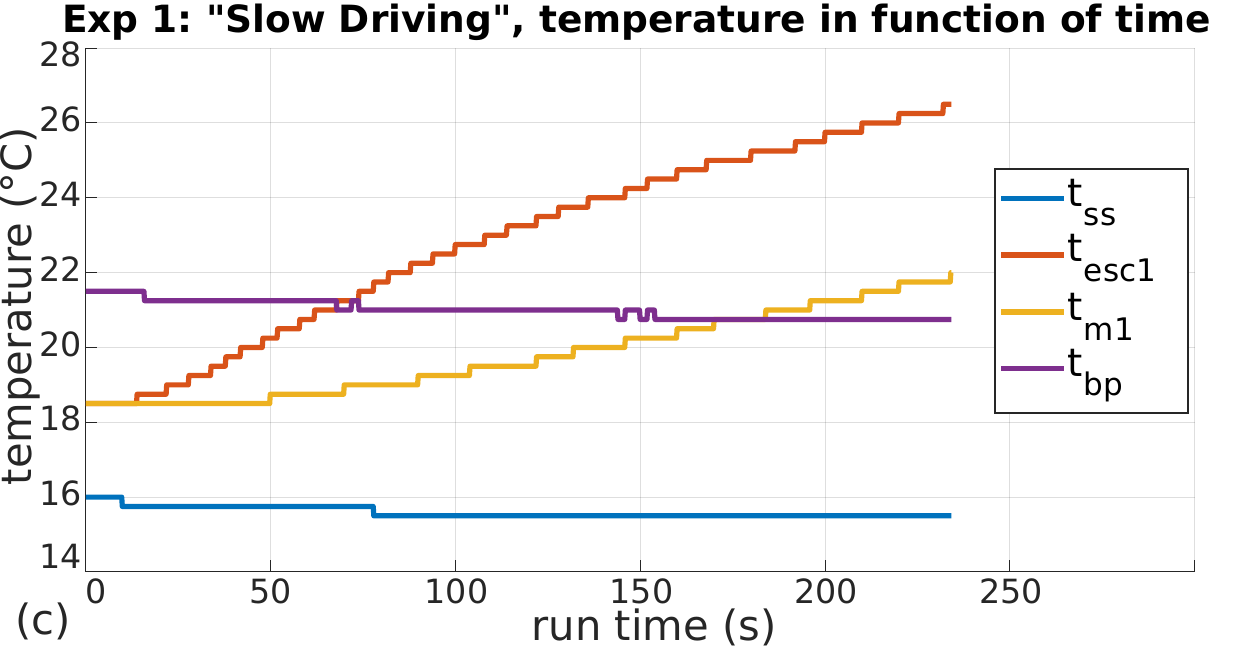}
	\end{subfigure}
	\begin{subfigure}{.49\textwidth}
		\centering
		\includegraphics[width=1\linewidth]{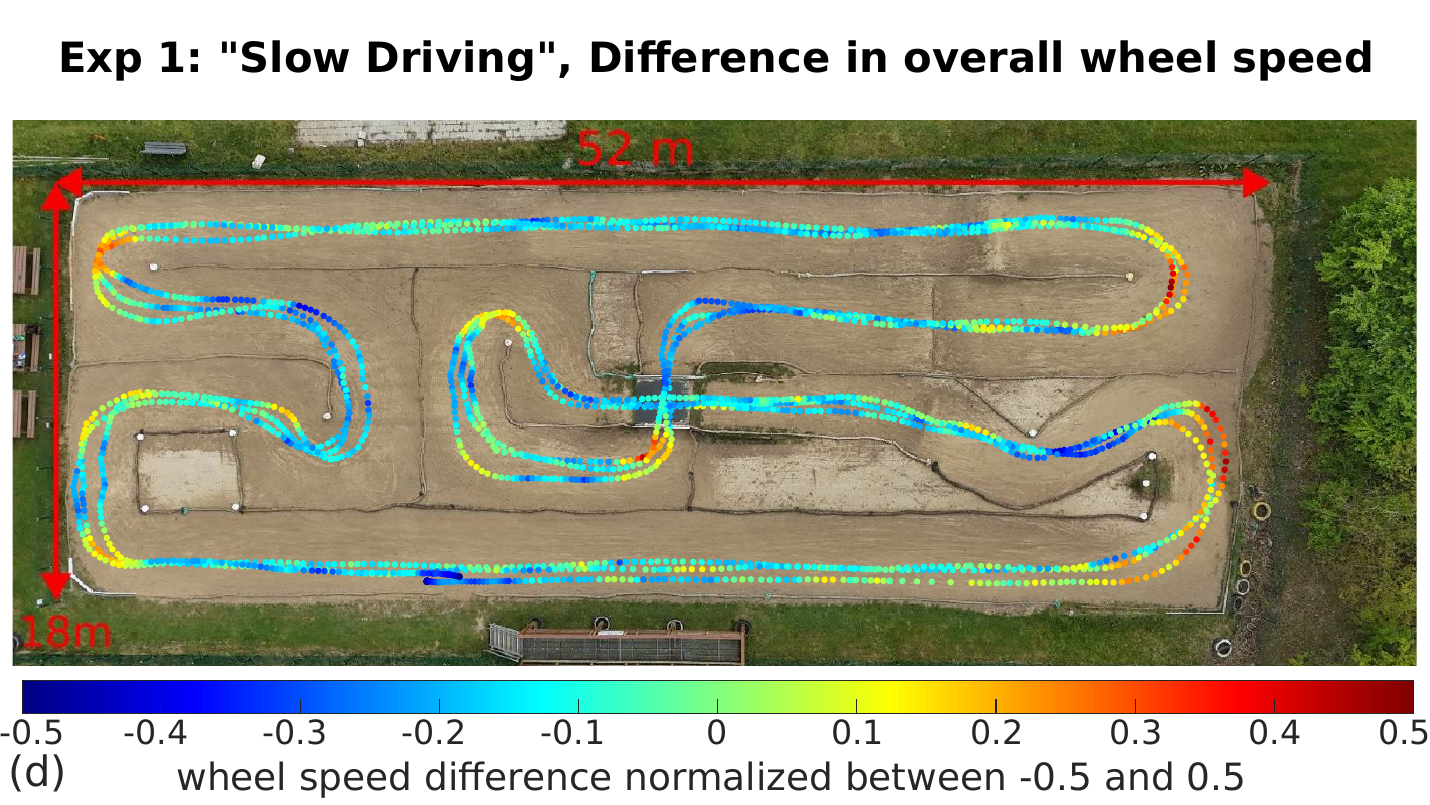}
	\end{subfigure}
	
	\caption{Results of experiment 1: \textit{Slow driving over the track}. Here, tree laps are driven. (a) representing the slow driven track \ac{GPS} data path. One can see that the data acquired form the three laps do indeed have a correlation in terms of \ac{GPS} position. Nevertheless, some small deviations are still visible due to the drivers chosen paths. (b) representing the pitch angle of three laps where similarities can be observed at the locations of bumps. Furthermore, a consistency of the sampled pitch data is present at similar bumps. Note that a positive pitch represents a downwards movement and a negative pitch, an upwards movement. (d) represent the difference between all wheels. It can be observed that in turnings, wheel speed differs due to the differentials in place. The used scale is normalized to give a clear difference.}	
	\label{fig:exp_1}
\end{figure*}

\begin{figure*}[t]
	\centering
	\begin{subfigure}{.49\textwidth}
		\centering
		\includegraphics[width=1\linewidth]{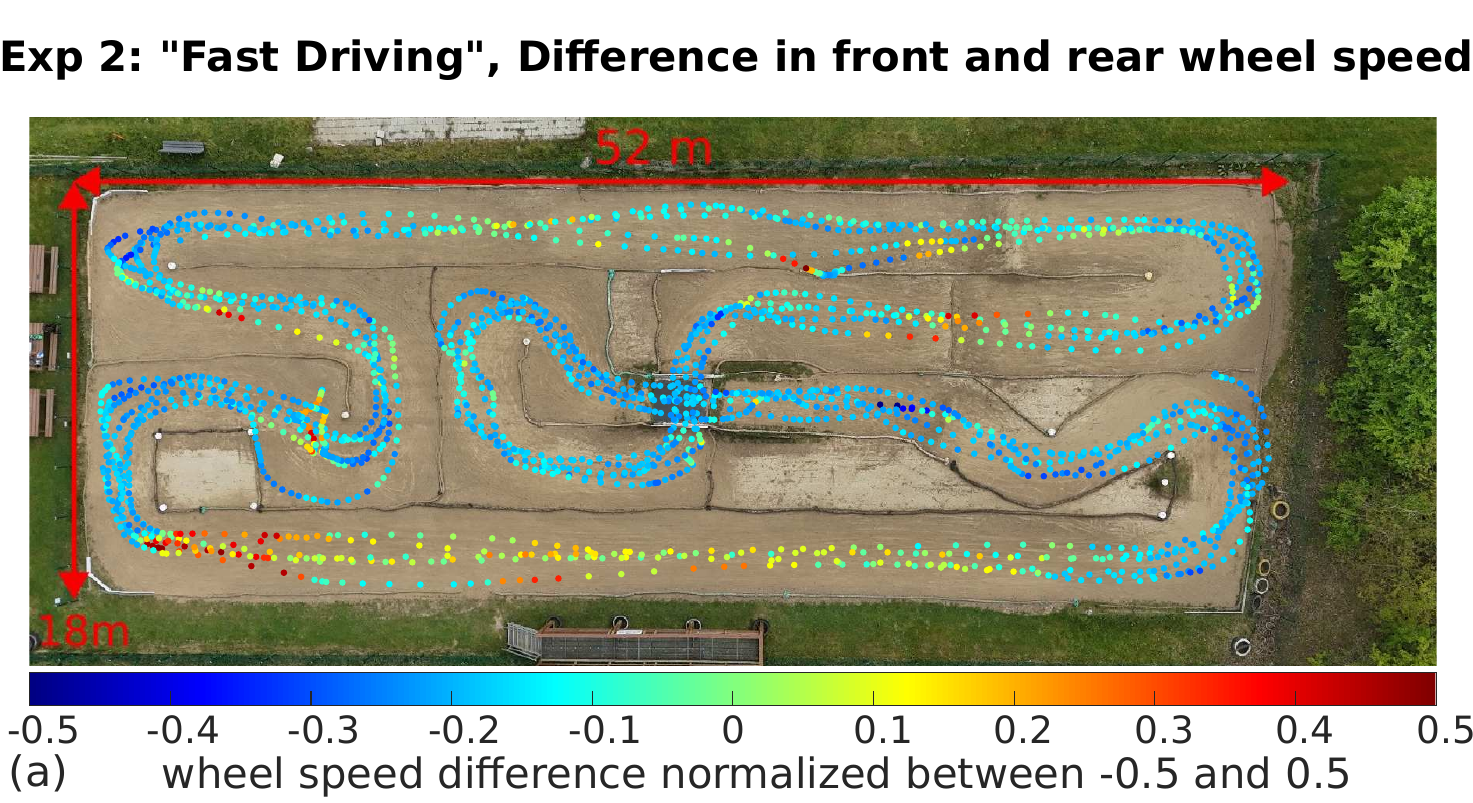}
		\vspace{0cm} % remove this line, and the image drops down for some reason
	\end{subfigure}
		\vspace{0.25cm}
	\begin{subfigure}{.49\textwidth}
		\centering
		\includegraphics[width=1\linewidth]{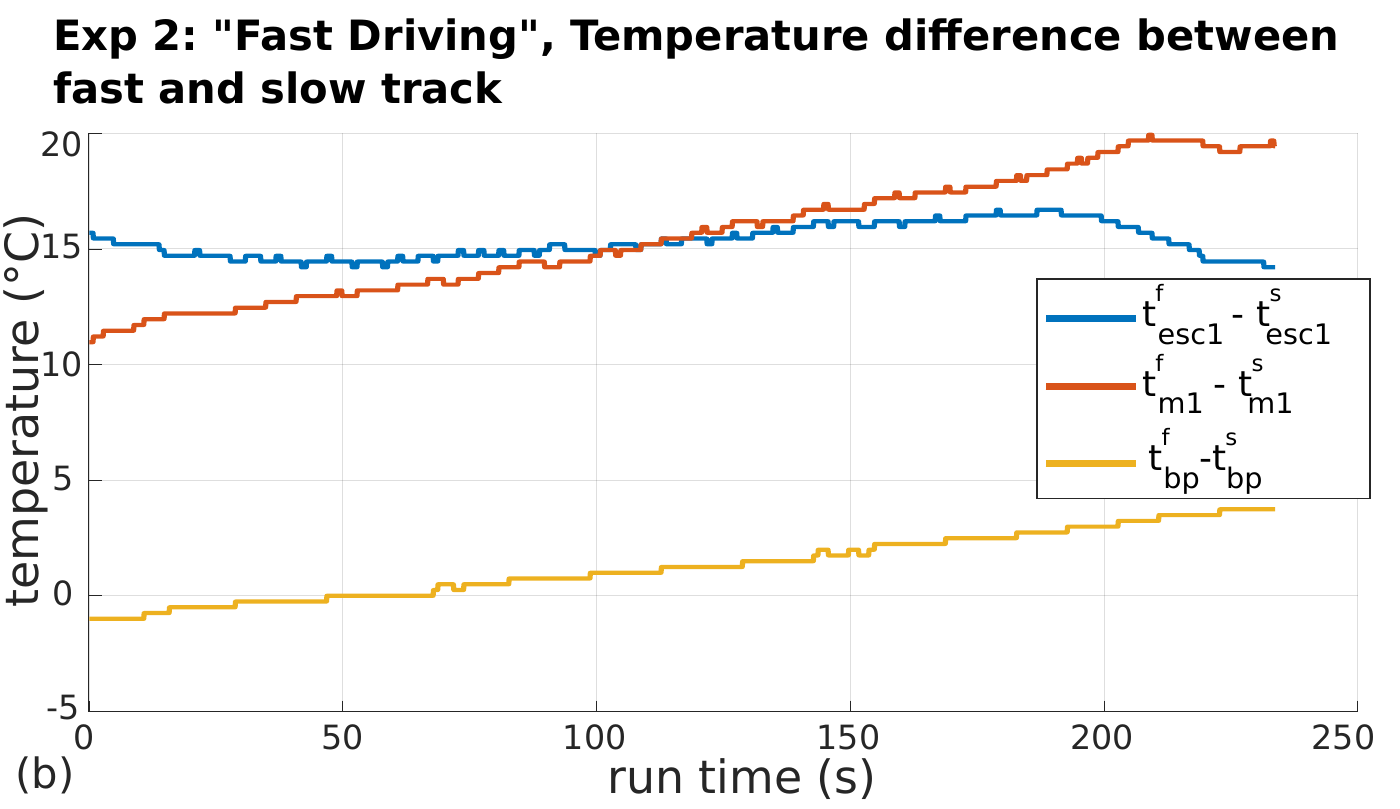}
	\end{subfigure}

	\begin{subfigure}{.49\textwidth}
		\centering
		\includegraphics[width=1\linewidth]{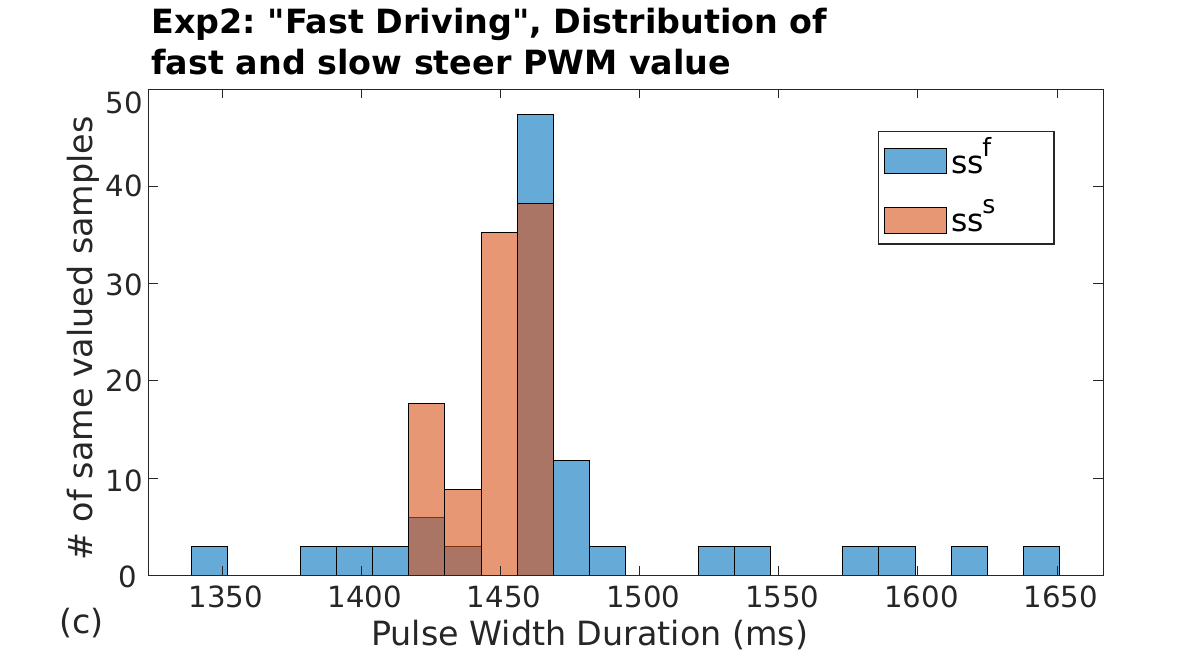}
		\label{fig:exp_2_dist_fast_slow_0}
	\end{subfigure}
	\begin{subfigure}{.49\textwidth}
		\centering
		\includegraphics[width=1\linewidth]{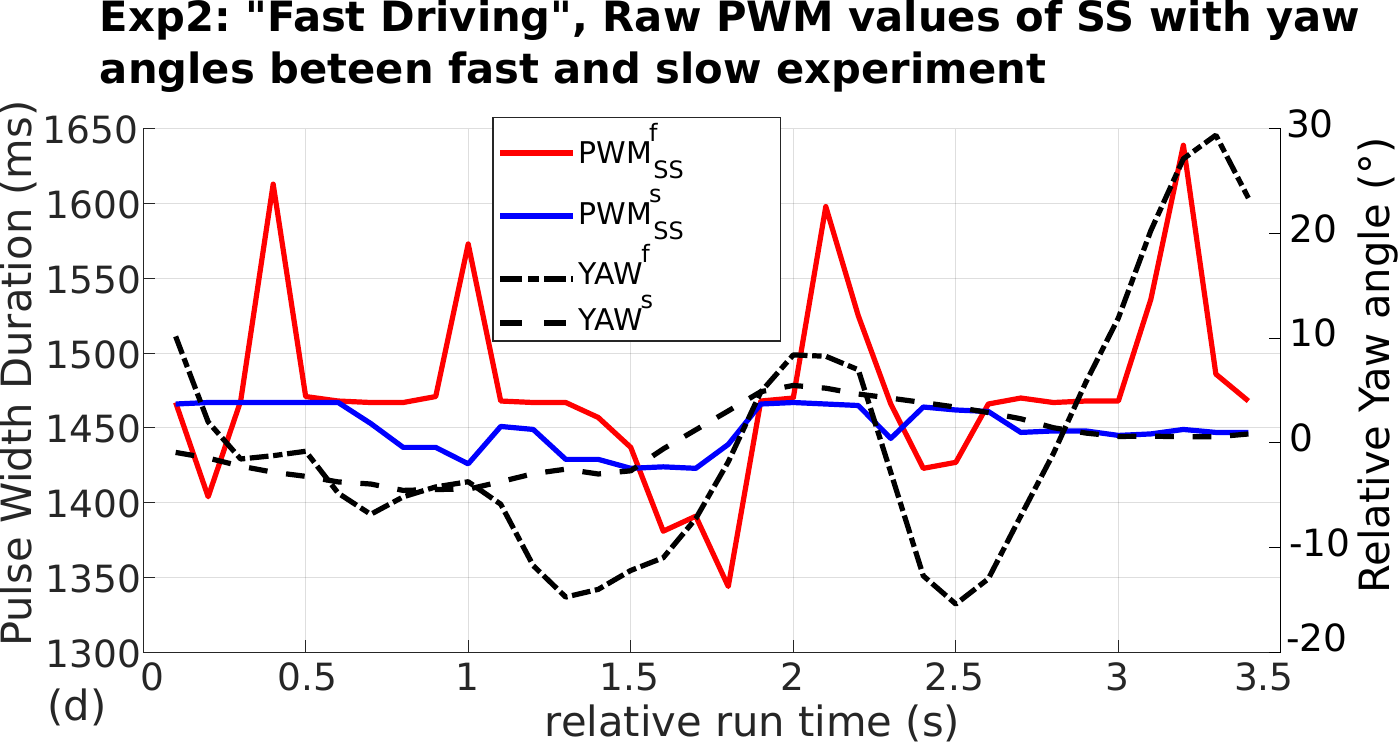}
		\vspace{0.0cm}
		\label{fig:exp_2_dist_fast_slow_1}
	\end{subfigure}
	\caption{Results of experiment 2: \textit{Fast driving over the track}. 
		(a) representing the wheel speed difference between front and rear. Furthermore, a positive difference reflects to faster rotating front wheels while a negative difference concludes to faster rotating rear wheels. One can observe that in the straight line, the front wheels rotate significant faster compared to the rear wheels. Similar behavior presents itself on hill features on the track. 
		(b) presents the temperature difference between experiment two and experiment one. Here, three temperature sources are used that is $t_{esc1}$, $t_{m1}$, and $t_{bp}$. One can see that $t^f_{m1} - t^s_{m1}$ and $t^f_{bp}-t^s_{bp}$ produce more heat over time where $t^f_{esc1}-t^s_{esc1}$ has similar heat generation over time. 
		(c) gives insights in the steer correction which was made during straight track paths. In particularly, the lowest straight path depicted in Fig. \ref{fig:predefined_path}. It is noticeable that in experiment two, more out-of-bound corrections where needed in order to maintain a straight line compared to experiment one. 
		(d) presents the Yaw angle for experiment one $YAW^s$ and two $YAW^f$ over time which, indeed, represents the actual vehicle movement as the counter steering will alter the Yaw angle for experiment one $PWM^s_{SS}$ and two $PWM^f_{SS}$. Yet again, it is visible that there is more Yaw motion followed by more steering motion to counter the Yaw motion.}	
	\label{fig:exp-2}
\end{figure*}

\begin{figure}
	\centering
	\begin{subfigure}{.48\textwidth}
		\centering
		\includegraphics[width=1\linewidth]{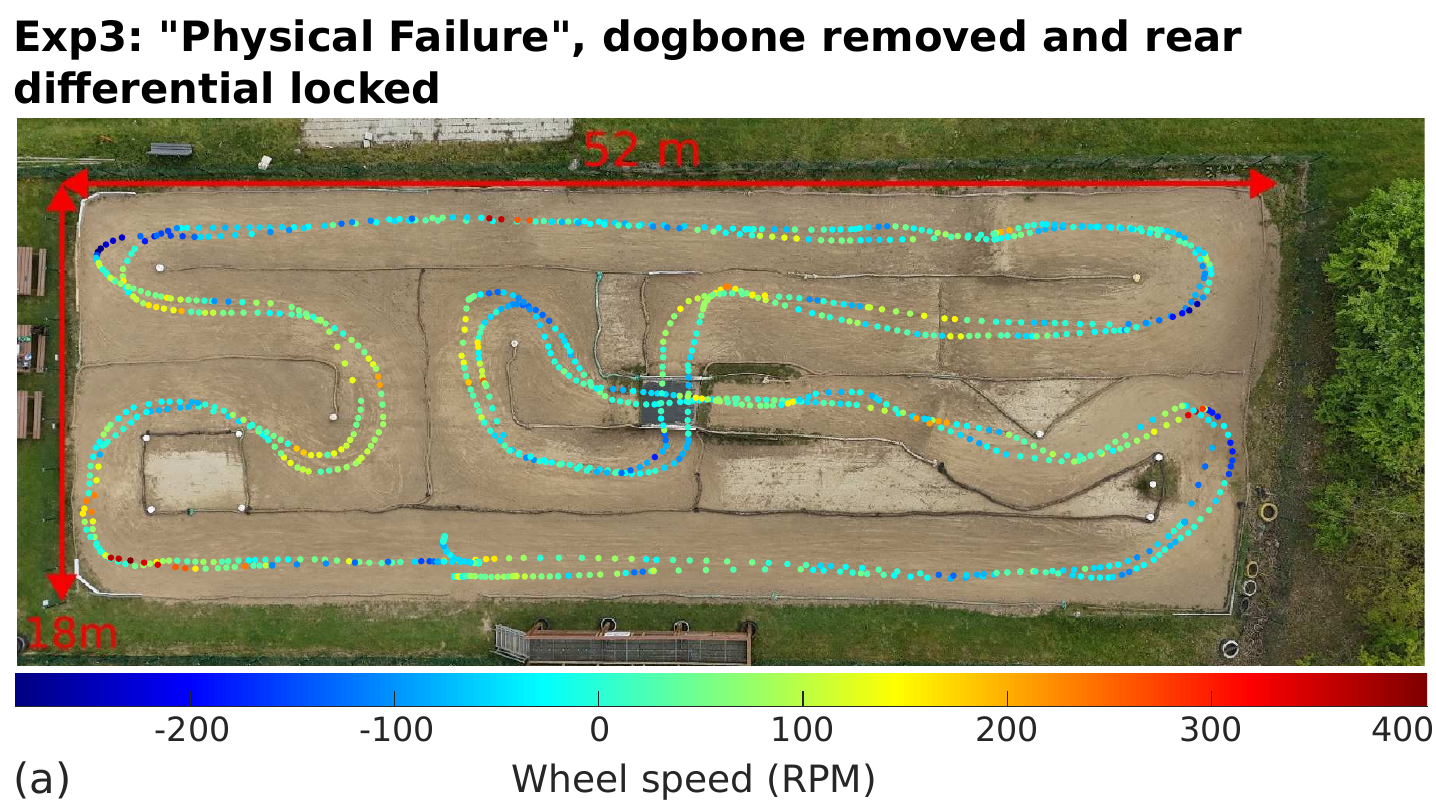}
	\end{subfigure}
	\begin{subfigure}{.48\textwidth}
		\centering
		\includegraphics[width=1\linewidth]{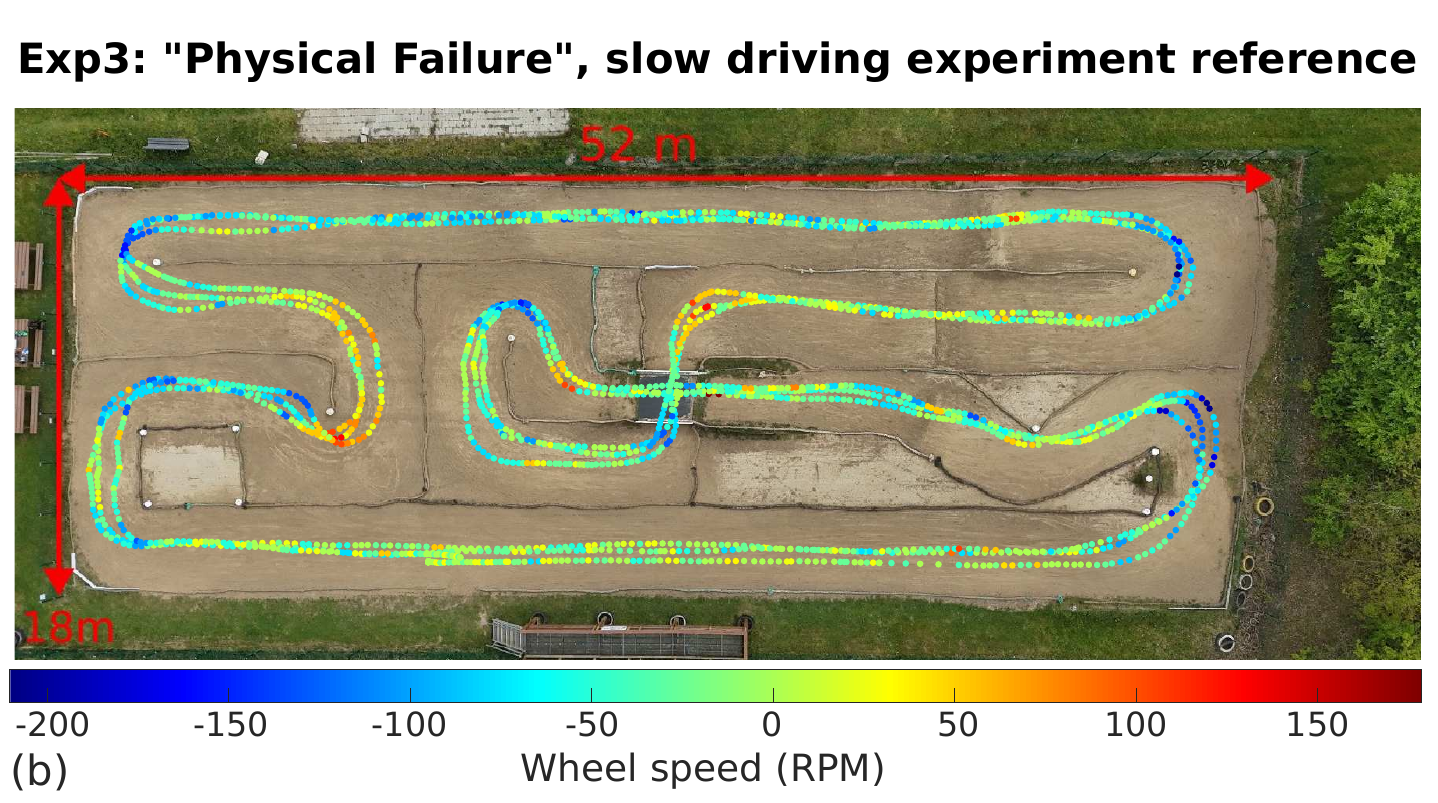}
	\end{subfigure}
	\caption{Presenting the difference in front and rear wheel speed (a) presenting wheel measurements from experiment three with a physical failure. (b) is a reference to \textit{Slow driving over the track} where no physical failure is present. One can see in (a) that the difference in wheel rotations is higher due to the fact that the rear differential is locked where the front is driven by the brushless motor. Moreover, (b) presents normal behavior where, indeed, the difference on straight paths is minimal. Moreover, the differential effects in cornering features are visible on both (a) and (b).
	}
		
	\label{fig:exp_3}
\end{figure}

\begin{figure}
	\centering
	\includegraphics[width=1\linewidth]{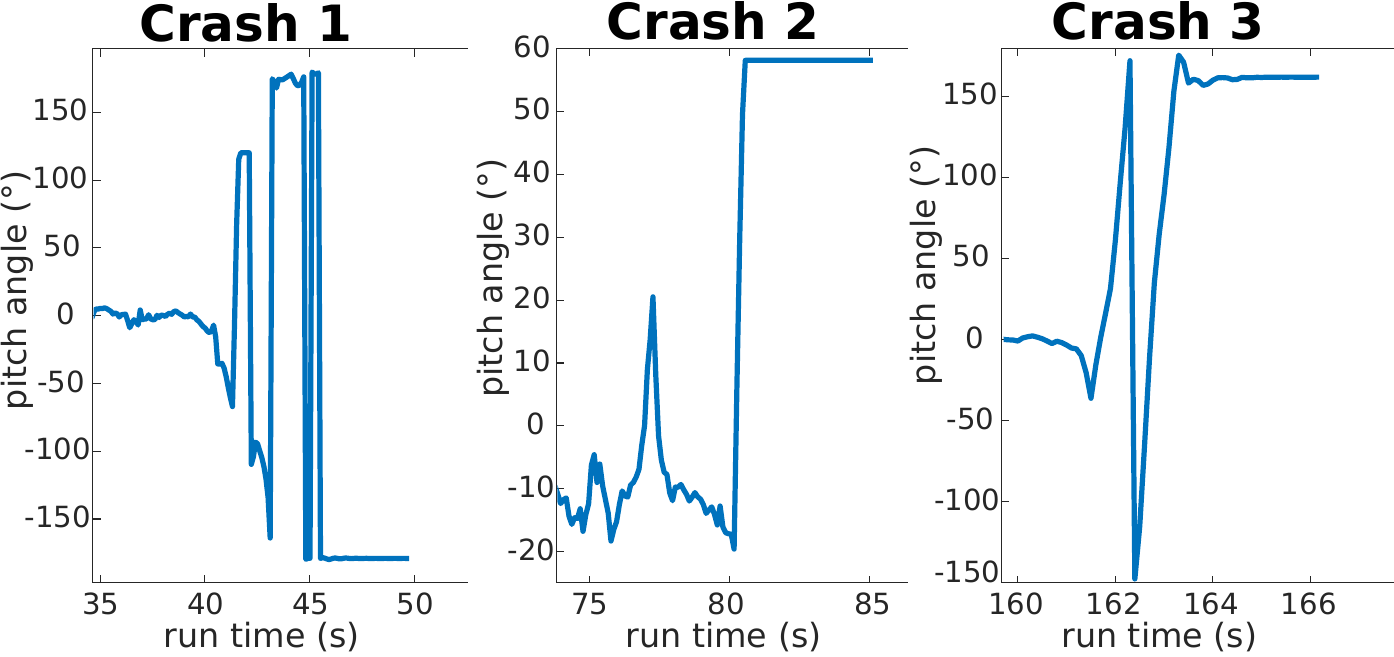}
	\caption{Presenting crash data where the pitch angle gives insight in how the crash evolved in term of tumbles around the y-axis. One can see that in \textit{Crash 1} and \textit{Crash 3}, thumbing is occurring due to the fact that the pitch-angle flips its axis multiple times in a short time-span. However, in \textit{Crash 2}, no thumbing is present resulting in a flat fall. In all cases, the \ac{RCV} ended on its roof.}
	\label{fig:exp4crashcombined}
\end{figure}

As previously discussed, the developed framework serves as a platform for verifying and validating \ac{ADAS} algorithms under challenging conditions. These algorithms rely on various measurements collected by the \ac{DAQ}. Therefore, the measurement system in place needs to be validated as well. Furthermore, these measurements must be consistent and reliable while maintaining a link to the real-world phenomenal. 
To achieve this, a predefined path will be driven multiple times in different experiments such as slow and fast driving. Moreover, the data collected during these runs will be analyzed to assess the accuracy and consistency of the \ac{DAQ}. Reliable measurements should yield strong correlations across comparable runs, demonstrating both stability and repeatability.
Additionally, to verify alignment with real-world behavior, specific and predictable vehicle actions will be performed, where outcomes can be anticipated. For instance, increasing throttle could result in higher wheel speed, and turning left should produce a measurable yaw change. These controlled tests confirm that the system's outputs accurately reflect the vehicle’s physical behavior.
The predefined path, shown in Figure~\ref{fig:predefined_path}, will be used for data collection. Furthermore, four experimental scenarios will be conducted on this track to evaluate the stability and accuracy of the \ac{DAQ}.

The first experiment named \textit{Slow driving over the track} will give insights in the overall \ac{DAQ} performance without stressing the physical frame. This gives the opportunity to test for example the accuracy of the used sensors. 
Experiment two, \textit{Fast driving over the track}, can give insights when the \ac{RCV} is performing an intensive task where bumps, dust, rocks and crashes present themselves more aggressively. Furthermore, in this experiment, it could be visible how the side-slip behaves in corners. Additionally, the manual correction a driver has to make on straight tracks, often conducted at high acceleration, could increase compared when the driver is driving the \ac{RCV} slowly over the track. 
Experiment three, \textit{Detach the rear wheel dogbones and lock the rear differential}, was conducted to determine whether a physical malfunction in the \ac{RCV} can be reliably detected from the collected data. 
Experiment four, \textit{Unintentional crashes}, is unique due to the fact that it is not forge-able and occurs when it is least expected. Nevertheless, these crashes are vital to test the durability of the whole \ac{DAQ}. During such a crash, often, a tremendous force on impact is present. This force could damage the structural integrity as well as the electronics connections with the framework. Not to mention that after the crash, the race is not over and the \ac{RCV} should be able to start where it landed. Therefore, the \ac{DAQ}, needs to be resilient enough for these occasions.

\subsection{Slow driving over the track}

During this experiment, the \ac{RCV} was driven slowly to minimize high accelerations and de-accelerations, resulting in 2337 samples at 10~\si{\hertz} over a total period of 233.70~\si{\second}. The \ac{GPS} data clearly illustrates the \ac{RCV} trajectory, with overlapping paths from the first three laps shown in Fig. \ref{fig:exp_1}a. Variations in terrain are reflected in the pitch angle changes illustrated in Fig. \ref{fig:exp_1}b, indicating localized slopes, dips, or ridges along the track \cite{el2023road}. These variations correspond to the physical terrain features present in Fig. \ref{fig:predefined_path}. 
Under these slow driving conditions, minimal acceleration and wheel slip result in lower power consumption and reduce heating. Consequently, the temperatures of steering servo $t_{ss}$, \ac{ESC} $t_{esc1}$, Brushless motor $t_{m1}$, and Li-Po battery pack $t_{bp}$ remained stable as shown in Fig. \ref{fig:exp_1}c.

The 1/8-scale \ac{RCV} features three differentials, one at the front, rear, and middle. Moreover, this enables wheels to rotate at different speeds during cornering. As illustrated in Fig. \ref{fig:exp_1}d, clockwise turns produce a positive wheel speed difference, while counterclockwise turns yield a negative wheel difference, consistent with differential mechanics \cite{el2023road}.

\subsection{Fast driving over the track}

This experiment was conducted under normal driving conditions, with the driver operating the \ac{RCV} at maximum performance. During these \textit{Fast driving over the track} scenarios, increased throttle leads to more frequent wheel slip. As shown in Fig. \ref {fig:exp-2}a, higher wheel slip is observed on straight sections and hills. Additionally, the increased speed results in wider spacing between \ac{GPS} sample points, compared to the \textit{Slow driving over the track} experiment.

While temperatures remained relative low in the \textit{Slow driving over the track} experiment, faster movement and higher current draw in this experiment leads to noticeable increase. Fig. \ref{fig:exp-2}b compares the temperature differences of the \ac{ESC} $t^f_{esc1} - t^s_{esc1}$, battery pack $t^f_{bp}-t^s_{bp}$, and brushless motor $t^f_{m1}-t^s_{m1}$ relative to experiment one. As expected, $t^f_{m1}-t^s_{m1}$  and $t^f_{bp}-t^s_{bp}$ increase more whereas $t^f_{esc1} - t^s_{esc1}$ remained relatively unchanged, likely due to the \ac{ESC}s active cooling system deployed onto the \ac{ESC}.

Speed plays a critical role on straight sections where the drivers aim is to close gaps with opponents. However, rapid acceleration can destabilize the \ac{RCV}. Consequently, compensation on steering input is needed to maintain a straight trajectory. It is observable in Fig. \ref{fig:exp-2}c and \ref{fig:exp-2}d that the driver applies more frequently and extremer steering corrections compared to the \textit{Slow driving over the track} experiment, indicating a higher degree of vehicle instability during high accelerations.

\subsection{Detach the rear wheel dogbones and lock the rear differential}

During off-road racing it is vital to detect structural and physical damage which can influence the performance of the \ac{RCV}. Furthermore, when these occasions occur, the driver must be aware of this in order to act fast such that he can drive the \ac{RCV} at its optimum. Therefore, it is vital to detect these failures or issues during races to make sure the \ac{RCV} is in optimal conditions during the race itself. Moreover, in this experiment \textit{Detach the rear wheel dogbones and lock the rear differential}, a failure is used that could occur. In this experiment, the rear wheels are free to spin while the rear differential is locked such that all motor force is driven to the front wheels. This phenomenon is visible in Fig. \ref{fig:exp_3}a where indeed a higher difference is observed indicating that the front wheel is rotating faster compared to experiment \textit{Slow driving over the track} depicted in Fig. \ref{fig:exp_3}b. Furthermore, A special phenomenon is occurring due to the fact that only one wheel of front and rear is taken for this data plot. Moreover, at cornering, differences in speed is present where one can see that for counter clock wise rotations, the difference is lesser where for clock wise rotations the difference is greater. Additionally, the arcs covered by the front and rear wheel are different in diameter which, indeed, give these phenomena.

\subsection{Unintentional crashes}

Crashes can occur unexpectedly and are typically unintentional and undesirable during races. Moreover, a crash can cause longer lap times and in worst case, significant damage to the \ac{RCV} with disastrous consequences \cite{app9235126}. However, crashes can also provide valuable insights on revealing how they occurred and informing future detection or prevention strategies \cite{mastinu2014road}. Additionally, pre-crash data can help identify mistakes or patterns that lead to such incidents. Therefore, to extract meaningful insights, it is essential that the \ac{DAQ} continues sampling data consistently during crashes, just as it does during normal operation. In Fig. \ref{fig:exp4crashcombined}, multiple crashes are presented through the pitch angle. Notably, the pitch angle flips one or more times, indicating that the \ac{RCV} is tumbling. This confirms that the \ac{DAQ} remains operational even under extreme mechanical stress.

\section{Discussion and Conclusion}

% Conclustion
This paper presents a \ac{DAQ} optimized for deployment on an off-road \ac{RCV}, designed for collecting consistent and representative data to be used in \ac{ADAS} development. The system enables monitoring of the \ac{RCV}s condition both locally or remotely during races. Furthermore, this data can be utilized for real-time behavioural optimization. Additionally, it helps identify potential hardware or physical failures before they occur. Moreover, during a race, key metrics such as driver speed, path selection, wheel slip, and side-slip interventions can be analysed and compared across races or vehicles to assess overall performance.

% Discussion
Experimental results confirm that the \ac{DAQ} accurately captures real-world behaviour, following vehicle motion in a consistent and repeatable manner. Integrated into a custom \ac{LLC}, the system demonstrated robustness, even during crash scenarios, while maintaining responsiveness during high-speed off-road operation. However, the current prototype has several limitations that is it can not measure individual motor speeds, \ac{ESC} and motor currents, or the true steering angle. It also lacks hardware redundancy, making it less suitable for safety-critical applications. Furthermore, as a prototype, it is not scalable across different platforms and lacks basic safety features such as reverse polarity protection and current limiting.

%Highlight future work
This work establishes a solid foundation for future student developments, providing a comprehensive platform for advancements and refinements. It serves as a starting point which enables students to build upon existing technologies while fostering the exploration of new ideas and innovative approaches. Rather than focusing on groundbreaking scientific discoveries, this project lays the foundation for a robust platform that holds the potential to make a substantial impact within the context of engineering education. One key enhancement is developing a custom, integrated \ac{PCB} that includes all current features along with additional safety circuits. Moreover, implementing \ac{ADAS} algorithms to support or replace the human driver remains an important direction. Ultimately, these systems could lead to fully autonomous \ac{RCV}s capable of navigating complex off-road environments in a race condition.

\section{Acknowledgment}
Gratitude is extended to Corally for providing the RC vehicle chassis used in this study. The robust construction and mechanical precision of the Corally platform offered a reliable foundation for the integration and validation of the \ac{DAQ} system. Furthermore, appreciation is also due to RC Racing Fun Bornem for granting access to their RC off-road track which served as a consistent and controlled environment for conducting the experimental trials. The support from both Corally and RC Racing Fun Bornem had significantly impact on the successful completion of this work.

\newpage

\bibliographystyle{IEEEtran}
\bibliography{../../Resources/bibliography}

\begin{acronym}
	\acro{ADAS}{Advanced Data Acquisition System}
	\acro{PPR}{Pulses Per Revolution}
	\acro{MCU}{Micro Controller Unit}
	\acro{RCV}{Remote Controlled Vehicle}
	\acro{DAQ}{Data Acquisition System}
	\acro{SDR}{Software Defined Radio}
	\acro{WAN}{Wide Area Network}
	\acro{LAN}{Local Area Network}
	\acro{PAN}{Personal Area Network}
	\acro{DSRC}{Dedicated Short Range Communication}
	\acro{BLE}{Bluetooth Low Energy}
	\acro{ISM}{Industrial, Scientific and Medical}
	\acro{UWB}{Ultra-Wideband}
	\acro{V2V}{Vehicle-to-Vehicle}
	\acro{V2I}{Vehicle-to-Internal}
	\acro{V2E}{Vehicle-to-Everything}
	\acro{ADC}{Analogue to Digital Converter}
	\acro{LPF}{Low Pass Filter}
	\acro{ANNs}{Artificial Neural Networks}
	\acro{RLS}{Recursive Least Squares}
	\acro{LKF}{Linear Kalman Filter}
	\acro{EKF}{Extended Kalman Filter}
	\acro{UKF}{Unscented Kalman Filter}
	\acro{TCS}{Traction Control System}
	\acro{ESC}{Electronic Speed Controller}
	\acro{ABS}{Anti-lock Braking System}
	\acro{RC}{Radio-Controlled}
	\acro{CIA}{Confidentiality, Integrity and Availability}
	\acro{MTTE}{Maximum Transmissible Torque Estimation}
	\acro{AFS}{Active Front Steering}
	\acro{DYC}{Direct Yaw-moment Control}
	\acro{PID}{Proportional Integral Derivative}
	\acro{IMU}{Inertial Measurement Unit}
	\acro{LLC}{Low Level Controller}
	\acro{HLC}{High Level Controller}
	\acro{UAV}{Unmanned Areal Vehicle}
	\acro{I2C}{Inter-Integrated Circuit}
	\acro{GPS}{Global Positioning System}
	\acro{AWD}{All Wheel Drive}
	\acro{FWD}{Front Wheel Drive}
	\acro{RWD}{Rear Wheel Drive}
	\acro{PCB}{Printed Circuit Board}
\end{acronym}
\end{document}